\documentclass{article}

\usepackage{arxiv}

\usepackage{threeparttable}
\usepackage{tabularx}
\usepackage{amsmath}
\usepackage{enumerate}

\usepackage[utf8]{inputenc} 
\usepackage[T1]{fontenc}    
\usepackage{hyperref}       
\usepackage{url}            
\usepackage{booktabs}       
\usepackage{amsfonts}       
\usepackage{nicefrac}       
\usepackage{microtype}      
\usepackage{cleveref}       
\usepackage{graphicx}
\usepackage{natbib}
\usepackage{doi}

\usepackage[group-separator={,},output-decimal-marker={.}]{siunitx}

\title{Posterior accuracy and calibration under misspecification in Bayesian generalized linear models}

\date{}

\author{
Maximilian Scholz \\
	Cluster of Excellence SimTech\\
	University of Stuttgart\\
 Germany \\
	\texttt{research.scholz@mailbox.org} \\
	\And
 Paul-Christian Bürkner \\
	Department of Statistics\\
 TU Dortmund University\\
 Germany\\
}

\renewcommand{\headeright}{}
\renewcommand{\shorttitle}{Parameter recovery under misspecification in GLMs}

\hypersetup{
pdftitle={Posterior accuracy and calibration under misspecification in Bayesian generalized linear models},
pdfauthor={Maximilian Scholz, Paul-Christian Bürkner},
pdfkeywords={generalized linear models, model misspecification, likelihood family, link function, posterior calibration, simulation study},
}

\begin{document}
\maketitle

\begin{abstract}
Generalized linear models (GLMs) are popular for data-analysis in almost all quantitative sciences, but the choice of likelihood family and link function is often difficult. This motivates the search for likelihoods and links that minimize the impact of potential misspecification.
We perform a large-scale simulation study on double-bounded and lower-bounded response data where we systematically vary both true and assumed likelihoods and links.
In contrast to previous studies, we also study posterior calibration and uncertainty metrics in addition to point-estimate accuracy.
Our results indicate that certain likelihoods and links can be remarkably robust to misspecification, performing almost on par with their respective true counterparts. Additionally, normal likelihood models with identity link (i.e., linear regression) often achieve calibration comparable to the more structurally faithful alternatives, at least in the studied scenarios.
On the basis of our findings, we provide practical suggestions for robust likelihood and link choices in GLMs.
\end{abstract}


\section{Introduction}
\label{sec:introduction}

Generalized linear models (GLMs) are a popular choice for data-analysis in almost all quantitative sciences, offering an easy to interpret additive structure and rich mathematical theory \citep{gelman_regression_2020, harrell_regression_2015, nelder_generalized_1972, gill_generalized_2001}. Despite their widespread use and abundance of teaching material, it remains a difficult task to build GLM-based regression models that are trustworthy, well-predicting, and well-explaining. 

GLMs consider a univariate response variable \(y\) that is is assumed to follow a parametric \emph{likelihood}, often called likelihood family \citep{bates_fitting_2015, burkner2017brms}, with one main centrality parameter $\mu$ that is predicted as well as zero or more auxiliary distributional parameters \(\psi_1, ..., \psi_P\) that are assumed constant over observations. For the \(n\textsuperscript{th}\) of a total of $N$ observations, we write
\begin{equation}
\label{GLM}
y_n \sim \mathrm{likelihood}(\mu_{n}, \psi_{1}, \ldots, \psi_{P}).
\end{equation}
The domain of all distributional parameters is specific to the given likelihood family. However, if the domain of $\mu$ does not span whole real line (e.g., if it can only take on positive values), a \emph{link function} has to be introduced such that
the transformed domain becomes unbounded. For the \(n\textsuperscript{th}\) observation we write
\begin{equation}
\label{link_gam}
\mathrm{link}(\mu_{n}) =  \sum_{j = 0}^J \beta_j f_j(X_n) \quad 	\Longleftrightarrow \quad
\mu_{n} = \mathrm{inv\_link} \left(\sum_{j = 0}^J \beta_j f_j(X_n) \right).
\end{equation}
In Equation \eqref{link_gam}, \(X_n\) denotes the vector \((x_{1n}, \ldots, x_{Kn})\) of predictor values of the \(n\textsuperscript{th}\) observation, \(f_j\) are deterministic transformations of the predictor variables and \(b_j\) are the regression coefficients.
The inverse link function $\mathrm{inv\_link}$ is also known as \emph{response function} as it transforms the unbounded linear predictor onto the possibly restricted domain of $\mu$.
A common special case of GLMs is \emph{linear regression}, using a normal likelihood and identity link function.
For an in-depth introduction to GLMs, we refer the reader to \cite{mccullagh2019generalized}.

From an applied analyst's perspective, there are four central design choices when implementing a GLM: (a) the likelihood family, (b) the link function, (c) the linear predictor term, and (d) whether and how to regularize (e.g., via informative priors or penalty terms; \cite{james_introduction_2013, gelman_bayesian_2013}).
All of these choices are mutually related \citep{gelman_prior_2017, gelman_bayesian_2020}, but specifically (a) and (b) are closely intertwined as the choice of link function depends on the support of $\mu$ and thus on the chosen likelihood. For example, it is common practice to use a logit link function for binary response data or the log link for positive response data \citep{mcelreath_statistical_2020, gelman_bayesian_2013}. Given the importance and practical relevance of choosing a likelihood family and link function, we focus on these two design choices in this paper.

In the following we assume that there exists an (unknown) true data-generating process (DGP) of the form presented in Equation \eqref{GLM}.
A GLM fitted on data that uses a different likelihood or link than the true DGP is misspecified, at least to some degree (see Section~\ref{sec:background} for a discussion of the related work).
Rather than proposing a process to choose an optimal likelihood and link for an opaque DGP, we are interested in the impact of misspecification (MS) itself.
This is motivated by what we perceive to be a lack of comprehensiveness in the literature in regard to the effect of MS on parameter recovery, that is, how accurately and precisely we can estimate the parameters of interest (here, the regression coefficients $\beta_j$).
Two aspects that we found particularly limiting were (1) the sole focus on point estimates, while ignoring estimation uncertainty (e.g., as measured by frequentist confidence intervals or Bayesian credible intervals; CIs) and (2) the general scope of simulation studies that tend to only compare small groups of likelihoods and very few links in limited simulation scenarios.
We approach this problem from a Bayesian perspective, as it allows us to implement and fit arbitrary GLMs without being limited to, say, likelihoods that come from the exponential family of distributions \cite{mccullagh2019generalized}.
However, we expect our findings to hold in frequentist settings as well, as explained in Section~\ref{sec:model-fitting}.

To that end, we present novel results of a large-scale simulation study,  consisting of over one million models, investigating the effect of both likelihood misspecification and link misspecification on posterior accuracy and calibration.
We also provide an overview of commonly used likelihood families and link functions in regression analyses across multiple research fields and offer practical advice based on our results.
For brevity, we will refer to posterior accuracy and calibration as \emph{parameter recovery}.
Concretely, our contributions are:

\begin{enumerate}[(i)]
    \item large-scale simulations of the effects of likelihood and link misspecification on Bayesian generalized-linear models for the most prominent likelihood families and link functions of double-bounded and lower-bounded data;
    \item the inclusion of posterior calibration and uncertainty metrics in addition to point-estimate accuracy to assess the effect of misspecification more comprehensively;
    \item advice for practitioners that arises from the presented results.
\end{enumerate}


\section{Related Work}
\label{sec:background}

A lot of foundational work on the effect of model misspecification on parameter recovery has been conducted in a maximum likelihood estimation (MLE) framework and is concerned with asymptotic properties, such as proofs of (quasi-)consistency and normality of the MLE under MS \cite[e.g.,][]{huber1967behavior, li1989regression, fahrmexr1990maximum, fomby2003maximum} and the development of statistical tests to detect MS \cite[e.g.,][]{white1982maximum, yu2019link, huang2021improved}.
While directly related to the present study, such asymptotic findings may not translate to finite, especially small, sample sizes.

More specifically on the topic of link misspecification, \cite{li1989regression} showed that many point estimators (e.g., MLE) are consistent up to a constant factor under certain regularity conditions. This showcases a notion of asymptotic robustness under link misspecification, but the practical implications are not immediately clear as the said constant is unknown in practice.
Further, there is a wealth of studies comparing MLE properties under link MS in binary regression comparing various potential link functions such as log, logit, and probit \citep{chen2018comparing}, logistic, Box-Cox, Cauchy, and Burr \citep{czado1992effect}, normal, Cauchy, Box–Cox, and Burr \citep{cangul2009testing}, logit, probit, and shifted-normal \citep{yu2019link}, as well as logit and generalized logit \citep{huang2021improved}. However, all aforementioned studies focus on point estimates and typically rely on asymptotic behaviour. In addition, it is a common occurrence to describe the effect of link MS only in terms of bias of the MLE. Due to the scale-dependency of parameter recovery in regression models, it is expected that link misspecification leads to bias as it changes the latent scale \citep{li1989regression}. A more meaningful measure of the effect of link MS on parameter recovery could be the calibration of null hypothesis significance tests (NHST; Type 1 and 2 error rates), at least for link functions that preserve zero as a common point of origin. We will explore this direction in the present simulation study.


Likelihood misspecification can be seen as a special case of link misspecification, so some of the general asymptotic results on point estimator robustness apply here as well \citep{li1989regression}. Focusing only on the likelihood,
a comparison between normal, log-normal and gamma likelihoods showed similar estimates for low variance data \citep{Dongen2007OnTD}. Similarly, \cite{jacqmin2007robustness} found that in linear mixed models point estimate and coverage rate of 95\% confidence intervals were robust to likelihood MS.
As the scope of these studies is limited, it remains unclear if these local behaviours generalize to other commonly used likelihoods. And, as with link MS, the focus is typically on point estimates, with analysis of calibration being largely lacking.

Another line of research is focused on detecting likelihood MS, rather than analysing robustness.
Many prominent approaches use goodness-of-fit or predictive performance metrics such as the ratio of maximum likelihoods \citep[e.g.][]{dumonceaux1973discrimination, Gupta2003DiscriminatingBW, vuong1989likelihood, lewis2011unified}, Akaike's information criterion (AIC) \citep[e.g.][]{dick2004beyond, ward2008review} and the Schwarz or Bayesian information criterion, deviance information criterion, and Bayes factors \citep{ward2008review}.
In addition, there is a large body of work on discriminating between distributions, another related problem. These studies tend to focus on pairwise or small-group comparisons, such as discriminating between a log-normal and Weibull distribution \citep{dumonceaux1973discrimination, kundu2004discriminating}, a Weibull and the generalized exponential distribution \citep{Gupta2003DiscriminatingBW}, an exponential--Poisson and gamma distribution \citep{barreto2015likelihood}, or between a log-normal, Weibull, and generalized exponential distribution \citep{dey2009discriminating}.
Similarly, there is work to detect link MS using MLE properties \citep[e.g.][]{yu2019link, huang2021improved}.
In comparison to misspecification detection, we deem the analysis of robustness more practically important, as some degree of MS is to be expected for most real-world problems \citep{buerkner_utility_2022}. And while tests identifying MS may contain information about its effect
\citep[e.g.,][]{scholz2023prediction}, that evidence is only indirect.


As highlighted earlier, and reflected here, most work on the effect of MS on parameter recovery has been done from an MLE perspective, focused on point estimates and oftentimes of asymptotic nature. The scope of existing studies focusing on finite sample behavior is limited and in many cases it is not clear how the results generalize across GLMs.
Thus, there is a need for a comprehensive, finite sample analysis of a wider range of commonly used likelihoods and links with the additional perspective of uncertainty calibration.

\section{Method}
\label{sec:method}

Assessing the impact of likelihood and link MS on parameter recovery requires knowledge of the true data-generating process (DGP). Additionally, our focus on small-sample-size behaviour and the consideration of GLMs with no available closed-form solutions makes an analytical approach infeasible. For these reasons, our analyses rely on large-scale simulations.

The results of this paper are part of a larger simulation study encompassing methods and metrics of predictive performance and parameter recovery. 
The results regarding predictive performance, along with a description of the general simulation setup, have been published in \cite{scholz2023prediction}.
To fit the aim of the present study, we deviated from the general simulation setup in \cite{scholz2023prediction} by removing the causally biased models and calculating appropriate model metrics, as further described below.

The simulation was implemented in R \citep{team2013r} using Stan \citep{stan_2022, carpenter2017stan} and brms \citep{burkner2017brms}. Our software packages bayesim \citep{scholz_bayesim_2022}, bayesfam \citep{scholz_bayesfam}, and bayeshear \citep{scholz_bayeshear} are available online.
A more thorough discussion of the included likelihoods and links, as well as all code and data are available in our online appendix \citep{online_appendix}.

\subsection{Likelihoods and Link Functions}
\label{sec:likelihoods_and_links}

The range of practically relevant likelihood classes is extensive and encompasses, among others, likelihoods for unbounded, lower-bounded, and double-bounded continuous data, as well as binary, categorical, ordinal, count, and compartmental data \citep{johnson_continuous_1995, johnson_discrete_2005, stasinopoulos_gamlss_2007, yee_vgam_2010, buerkner_bayesian_2021}. Here, we focus our efforts on GLMs with lower-bounded or double-bounded continuous likelihoods. Within these classes, we not only have several qualitatively different (non-nested) likelihood options, but can also study both main classes of non-identity link functions (i.e. single-bounded and double-bounded).

We used the same inclusion criteria as \cite{scholz2023prediction} and provide a short overview of the likelihood and link functions considered here.

\subsubsection{Likelihoods and links for double-bounded responses}
\begin{figure}[tb] 
\centering
\includegraphics[width=0.99\linewidth]{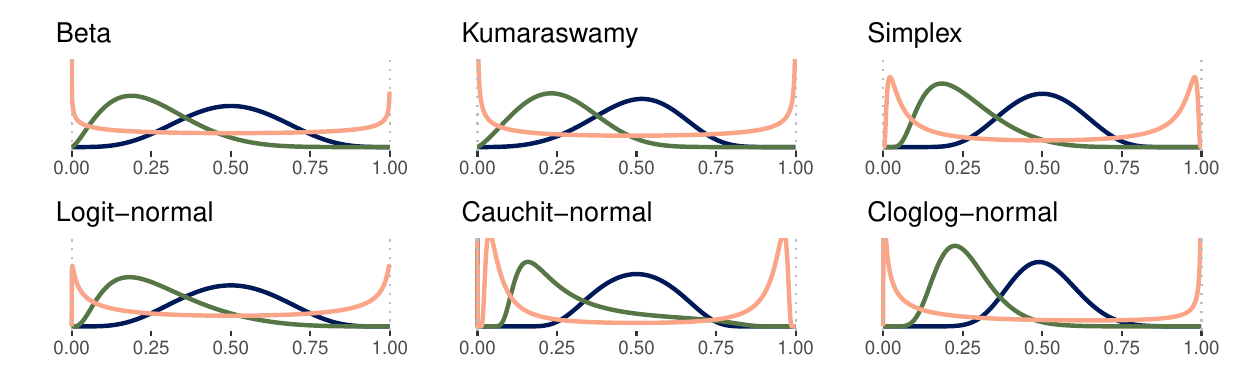}
\caption{Example illustrations of all included double-bounded densities each with three different shapes. The y-axis is truncated at 5 from above for better visibility of the different shapes.} 
\label{fig:unit_interval_densities}
\end{figure}

Without loss of generality, any double-bounded response $y$ with bounds $a, b$ can be linearly transformed to the unit interval by the transformation $f(y) = \frac{y - a}{b - a}$. Accordingly, it is sufficient to focus on likelihoods for unit interval data.
We included the beta \citep{espinheira_beta_2008}, Kumaraswamy \citep{kumaraswamy_generalized_1980}, simplex \citep{barndorff-nielsen_parametric_1991}, and transformed-normal \citep{atchison_logistic-normal_1980, kim_sample_2017} likelihoods.
The transformed-normal likelihoods arise from applying the link functions to the response variable $y$ (e.g., resulting in the logit-normal likelihood) instead of the location parameter $\mu$ as usual in standard GLMs. All of these likelihoods have two distributional parameters, one location (mean or median) and one scale (aka. shape) parameter.
Figure~\ref{fig:unit_interval_densities} shows prototypical densities for each likelihood, illustrating qualitatively different shapes they can accommodate.
The three distinct shapes are unimodal symmetric and asymmetric shapes as well as a bimodal (aka. bathtub) shape.
%
As link functions, we included the logit, cloglog, and cauchit links, each of them having qualitatively different properties \citep{yin_skewed_2020, jiang_new_2013, lemonte_new_2018, damisa_comparison_2017, fahrmeir_multivariate_1994, gill_generalized_2001, powers2008statistical, morgan_note_1992, koenker_parametric_2009, lemonte_new_2018}.
The logit link is based on the symmetric, light-tailed logistic distribution, the cloglog link is based on the asymmetric Gumbel distribution, while the cauchit link is based on the symmetric, heavy-tailed Cauchy distribution.
Since logit and probit yield almost indistinguishable results due to the similar shapes of the logistic and the normal distributions \citep{fahrmeir_multivariate_1994, gill_generalized_2001, powers2008statistical}, we decided against including the probit link despite its prominence.

\subsubsection{Likelihoods and links for lower-bounded responses}
\begin{figure}[tb] 
\centering
\includegraphics[width=0.99\linewidth]{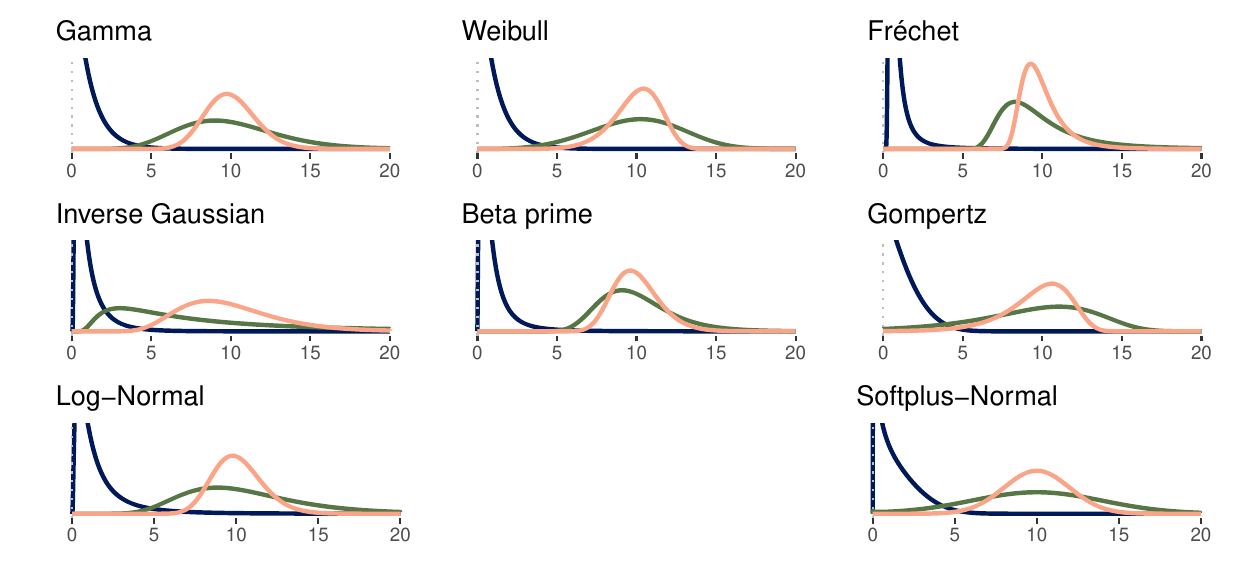}
\caption{Example illustrations of all included lower-bounded densities, each with three different shapes. The y-axis is truncated at 0.4 from above for better visibility of the different shapes.} 
\label{fig:lower_bounded_densities}
\end{figure}

Without loss of generality, any continuous response $y$ with lower bound $a$ can be linearly transformed to have a lower bound of zero by the transformation $f(y) = y - a$. Accordingly, it is sufficient to focus on likelihoods for strictly positive data.
We included the gamma, Weibull, Fréchet, beta prime, Gompertz, and transformed-normal likelihoods
all of which have two distributional parameters, namely location (mean or median) and scale or shape.
Figure~\ref{fig:lower_bounded_densities} shows example densities for each likelihood, illustrating qualitatively different kinds of shapes they can accommodate.
The three distinct shapes are unimodal thin tail and heavy tail shapes as well as a ramp shape. 
As link functions, we included the log and the softplus link. In contrast to the multiplicative log link,
softplus approaches the identity for larger values, thus approximating additive behavior of regression terms while enforcing positive predictions \citep{zheng2015improving, dugas2000incorporating}.

\subsubsection{Linear regression}
\label{sec:linear-regression}

In a nutshell, there are two kinds of approaches to choosing a likelihood.
The first is to search within the space of \emph{structurally faithful} \citep{buerkner_utility_2022} likelihoods that respect the variable type of $y$, for example, an exponential or Gamma likelihood for positive continuous data that has no or no known upper bound.
The second approach is just using a normal likelihood with identity link (i.e., linear regression) regardless of response type.
The latter approach is openly advocated for comparably rarely \citep{hellevik_linear_2009} but de-facto applied across many disciplines because of its convenience and interpretability of the obtained regression coefficients.
Still, there are obvious drawbacks of the "linear regression for all" approach.
For instance, it can produce predictions that are impossible in reality (e.g., negative counts). What is more, it may drastically distort effect size estimates, their uncertainty, and sometimes even their sign in certain cases \citep{stroup2015rethinking, martin_outgrowing_2017, williams_rethinking_2017, liddell_analyzing_2018}.
To investigate this topic further, we also fit models using a normal likelihood and identity link in our simulations.
Additionally, as an in-between approach between linear regression and structural faithful GLMs, we include the normal likelihood in combination with the appropriate (double- or lower-bounded) link functions.

\subsection{Data Generation}
\label{sec:data-generation}

\begin{figure}[tb]
\centering
\includegraphics[width=0.6\linewidth]{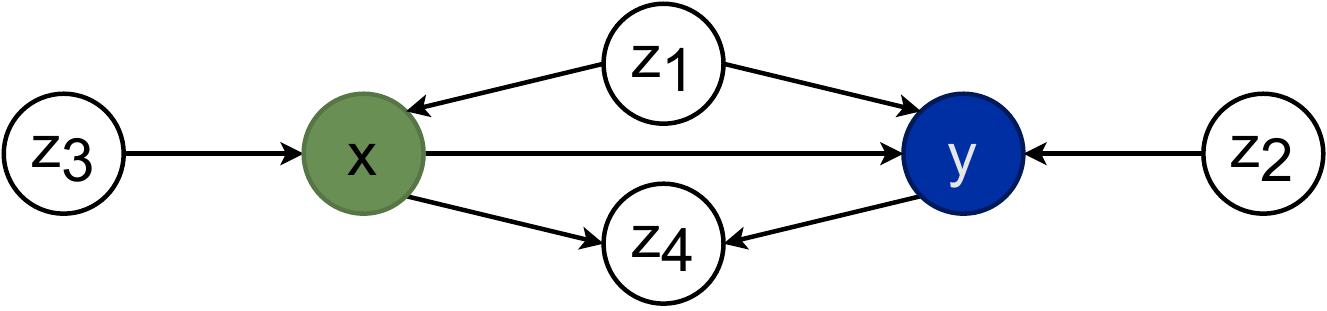}
\caption{Full data-generating graph. The ideal model is the subset of the full model that optimally estimates $\beta_{xy}$. Misspecifying with respect to each of the $z_n$ variables leads to four additional models, where excluding $z_1$ or including $z_4$ increases bias, while excluding $z_2$ or including $z_3$ increases variance of the estimation. As we don't want estimation bias caused by causal misspecification in our models, we only include models misspecifying $z_2$ and $z_3$.} 
\label{fig:dag}
\end{figure}

For each of the combinations of likelihood family, likelihood shape, and link presented in Section \ref{sec:likelihoods_and_links}, we simulated data sets based on a single, prototypical causal directed acyclic graph \citep[DAG;][]{pearl2009causality}.
The full DAG, shown in Figure~\ref{fig:dag}, consists of an outcome $y$, a treatment $x$, and four additional variables $z_1, z_2, z_3, z_4$ that correspond to four common, qualitatively different types of controls \citep{cinelli_crash_2020}. With respect to the effect of $x$ on $y$, $z_1$ is a fork, $z_2$ is an ancestor of $y$, $z_3$ is an ancestor of $x$, and $z_4$ is a collider. These archetypes, plus the pipe $(x \rightarrow z \rightarrow y)$ that behaves like a fork, represent the majority of controls occurring in reality \citep{cinelli_crash_2020}.
We then used the following model to simulate the data sets:
\begin{align*}
z_1 & \sim \mathrm{normal}(0, \sigma_{z_1}), \quad
z_2 \sim \mathrm{normal}(0, \sigma_{z_2}), \quad
z_3 \sim \mathrm{normal}(0, \sigma_{z_3}) \\
x & \sim \mathrm{normal}(\beta_{z_1x}z_1 + \beta_{z_3x}z_3, \sigma_x) \\
y & \sim \textrm{likelihood}(\text{inv\_link}(\alpha_y + \beta_{xy}x + \beta_{z_1y}z_1 + \beta_{z_2y}z_2), \phi) \\
z_4 & \sim \mathrm{normal}(\beta_{xz_4}x + \beta_{yz_4}y, \sigma_{z_4}).
\end{align*}
Here, $\phi$ denotes the second distributional (scale or shape) parameter of the specified likelihood family and $\text{inv\_link}$ denotes the inverse link (aka response) function. We chose normal distributions to generate all data variables except for $y$ to control the scope of our simulations. Each simulated data set contained $100$ observations.
Since the fitted models only have between 4 and 6 parameters, they are simple enough to be well identified on the basis of $100$ observations alone.
%
A general challenge for Bayesian simulations that use prior distributions over parameters is the tendency to produce extreme data sets \citep{gabry_visualization_2019, mikkola_prior_2021}.
Additionally, the uncertainty propagation from the prior can blur the distinctiveness of the different likelihood shapes.
To preserve control over the true DGP and to allow to control for the likelihood shapes during analysis, the parameters of the true DGP were thus set to fixed values, rather than being drawn from a prior distribution \citep{talts_validating_2020}.
Each likelihood's second distributional parameter $\phi$ as well as the intercept $\alpha_y$ were chosen to produce the likelihood shapes presented in Section~\ref{sec:likelihoods_and_links}. 
The individual coefficients for $x$ and the $z_i$ were calibrated so that the parameter recovery was imperfect for the ideal model while also preventing the causally misspecified models from consistently failing (see also Section \ref{sec:model-fitting}). The true causal effect $\beta_{xy}$ of $x$ on $y$ was either fixed to zero or set to a non-zero value that was calibrated together with all other coefficients.
To prevent response values from under- or overflowing to the lower- or upper boundaries numerically, we truncated them near the boundaries with a tolerable  error bound of $\varepsilon = 10^{-6}$. 

For each data generation configuration implied by fully crossing the design factors (see Table \ref{tab:data-gen-conf}), we generated $200$ data sets, which resulted in $14,400$ data sets each for the double- and single-bounded scenarios.

\begin{table}[tb]
\caption{Overview of data generation configurations.}
  \centering
    \begin{tabularx}{\linewidth}{ l X }
    \hline 
    \textbf{Factor} & \textbf{Levels} \\
    \hline 
    Double-bounded likelihoods & beta, Kumaraswamy, simplex, transformed-normal \\ 
    Double-bounded links & logit, cauchit, cloglog \\ 
    Double-bounded likelihood shapes & symmetric, asymmetric, bathtub \vspace{0.5em}\\ 
    Lower-bounded likelihoods & gamma, Weibull, transformed-normal, Fréchet, beta-prime, Gompertz \\ 
    Lower-bounded links & log, softplus \\ 
    Lower-bounded likelihood shapes & ramp, heavy tail, thin tail \vspace{0.5em} \\ 
    True $\beta_{xy}$ & zero, positive \\ \hline
    \end{tabularx}%
  \label{tab:data-gen-conf}%
\end{table}%

\subsection{Model Fitting}
\label{sec:model-fitting}

On each simulated data set, we fitted all models resulting from the fully crossed combination of likelihoods and links (see Section \ref{sec:likelihoods_and_links}) as well as a model with a normal likelihood and identity link to serve as a baseline (see Section \ref{sec:linear-regression}).
We then fit each resulting combination of likelihood and link with the three causally unbiased linear predictor terms implied by the DGP from Section~\ref{sec:data-generation} (see Table~\ref{tab:model-fit-conf}).
Here, we only include causally misspecified models that don't asymptotically bias the estimation of $\beta_{xy}$, as we are not investigating causal bias in this study. Thus remain the wrongful exclusion of $z_2$ or inclusion of $z_3$ to the ideal model, both increasing posterior variance (reducing precision) \citep{cinelli_crash_2020}.
In reference to R formula syntax, we will also use the term 'formulas' to refer to the various linear predictor terms in the following.
An overview of the model fit configurations is given in Table \ref{tab:model-fit-conf}.


The fully-crossed design results in $48$ fit configurations for the double-bounded models and $45$ fit configurations for the lower-bounded models.
Multiplied with $\num{14400}$ data sets each, this leads to a total of $\num{691200}$ double-bounded models and $\num{648000}$ single-bounded models, for a total of $\num{1296000}$ models fitted in our simulations.

\begin{table}[tb]
  \centering
  \caption{Overview of model fit configurations.}
    \begin{tabularx}{\linewidth}{ l X }
    \hline
    Factor & Levels \\
    \hline 
    Double-bounded Likelihoods & beta, Kumaraswamy, simplex, transformed-normal, normal \\ 
    Double-bounded Links & logit, cauchit, cloglog, identity (only with normal)\vspace{1em} \\ 
    
    Single-bounded Likelihoods & gamma, Weibull, transformed-normal, Fréchet, beta-prime, Gompertz, normal \\ 
    Single-bounded Links & log, softplus, identity (only with normal \\ 
    Formulas (right-hand side) & $x + z_1 + z_2$, $\; x + z_1$, $\; x + z_1 + z_2 + z_3$\\ \hline
    \end{tabularx}%
  \label{tab:model-fit-conf}%
\end{table}%

Contrary to what we would recommend in practical applications of Bayesian models, we used flat priors for all model parameters, as it is not clear to us how one would specify equivalent priors for the different auxiliary parameters $\phi$ across all considered likelihoods.
What is more, different links imply different latent scales, which render the regression coefficients' scales incomparable across (assumed) links and thus further complicate equivalent prior specification. 
In a real-world analysis, we would prefer to use at least weakly-informative priors \citep{stan_2022, gelman_bayesian_2013, mcelreath_statistical_2020}.
In pilot experiments (not reported here), we have confirmed that the differences in posteriors as well as the implied prediction metrics between models with flat vs. weakly informative priors are minimal for the models under investigation.
We argue that such minimal differences do not justify extensive evaluation of different prior choices, since this is not the focus of the present paper.
Additionally, the use of flat priors results in model estimations very similar to maximum likelihood estimation, which is why we expect the results of this study to generalize to frequentist models as well.  

All models were fitted using Stan \citep{carpenter2017stan, stan_2022} via brms \citep{burkner2017brms} with two chains, 500 warmup- and 2000 post-warmup samples, which resulted in $4000$ total post-warmup posterior samples per model.
We used an initialization range of $0.1$ around the origin on the unconstrained parameter space to avoid occasional initialization failures.
For all other MCMC hyperparameters, we applied the brms defaults \citep{burkner2017brms}.

\subsection{Model-Based Metrics}
\label{sec:model-based-metrics}

To measure parameter recovery of each fitted model, we used multiple metrics as detailed below. Implementations of these metrics are provided in the R packages posterior \citep{burkner2022posteriorR}, bayesim \citep{scholz_bayesim_2022}, and bayeshear \citep{scholz_bayeshear}.

We calculated the posterior bias and $\mathrm{RMSE}$ of the model's estimation of $\beta_{xy}$.
Given a true parameter value $\theta^\star$ and a set of $S$ corresponding posterior samples $\{ \theta^{(s)} \} := \{ \theta^{(s)} \}_{s=1}^S$, we compute the sampling-based posterior bias and $\mathrm{RMSE}$  as
\begin{equation}
\label{eq:bias}
\mathrm{bias}(\theta) := \mathrm{bias}(\{ \theta^{(s)} \}, \theta^\star) := \frac{1}{S} \sum^S_{s=1}\left( \theta^{(s)} \right)  - \theta^\star,
\end{equation}
 \begin{equation}
\label{eq:rmse_s}
\mathrm{RMSE}(\theta) := \mathrm{RMSE}(\{\theta^{(s)}\}, \theta^\star) := \sqrt{ \frac{1}{S} \sum^S_{s=1} \left(\theta^{(s)}  - \theta^\star \right)^2 }= \sqrt{\mathrm{bias}(\theta)^2 + \mathrm{Var} \left( \theta \right)},
\end{equation}
where $\mathrm{Var}(\theta) := \mathrm{Var}(\{ \theta^{(s)} \})$ denotes the variance over the posterior samples. Our analysis focuses on the true effect $\beta_{xy}$ and hence all metrics were computed for $\theta = \beta_{xy}$.
Furthermore, as we are interested in the magnitude of the bias but not in its direction, we use the absolute bias in our results. 

The above are reasonable measures for comparing models only if the assumed link coincides with the true link of the DGP. This is because the link determines the scale of the linear predictor and thus the comparability of the posterior samples $\{ \beta_{xy}^{(s)} \}$ with the true parameter value $\beta^\star_{xy}$. 
To enable a comparison of models using different links, we also calculated the false positive rate (FPR; i.e., Type I-error rate) and the true positive rate (TPR; i.e., statistical power; inverse of the Type II-error rate) implied by the central 95\% credible interval of $\{ \beta_{xy}^{(s)} \}$.
Significant CIs (no zero-overlap) count as true-positives in the case of non-zero true coefficients and false-positives in the case of a true coefficient of zero.
These metrics can be inferred from our simulations, because the true effect $\beta_{xy}$ was set to zero in some conditions (to study FPR) and to non-zero values in others (to study TPR).
In addition, we also present FPR and TPR in their combined form via receiver operating characteristic (ROC) curves \citep{zweig1993receiver} and the area under the ROC curve (AUC) \citep{bradley1997use}.

\section{Results}
\label{sec:results}

Below, for brevity, we focus on a subset of results from our simulations that represent the main overarching patterns we observed.
For example, the results for $| \mathrm{bias}(\beta_{xy}) |$ generally showed the same patterns as those for $\mathrm{RMSE}(\beta_{xy})$, such that we only present the latter here.
The complete results are available in our online appendix \citep{online_appendix}, together with the simulation data and code.

We excluded around 30000 models (2.3\% of all models) from the analysis as they did not converge. Around 25,000 of the non-converged models were fit with a Fréchet likelihood and softplus link, a Gompertz likelihood, or a cloglog link, probably due to numerical instabilities. 
Specifically, we treated  models as converged if the posterior samples $\{ \beta_{xy}^{(s)} \}$ yielded $\widehat{R} < 1.01$ and $\mathrm{ESS} > 400$ (for details on these thresholds, see \cite{vehtari_rank-normalization_2021}). Additionally, we required models to have less than $10$ divergent transitions out of a total of 4000 post-warmup iterations.
While in practice, we would like all models to converge with no divergent transitions, this would have required extensive manual intervention to resolve all individual sampling problems, which is practically infeasible in our large simulation setup of more than one million models in total.

\subsection{Double-bounded Results}

\begin{figure}[tb]
\centering
\includegraphics[width=\linewidth]{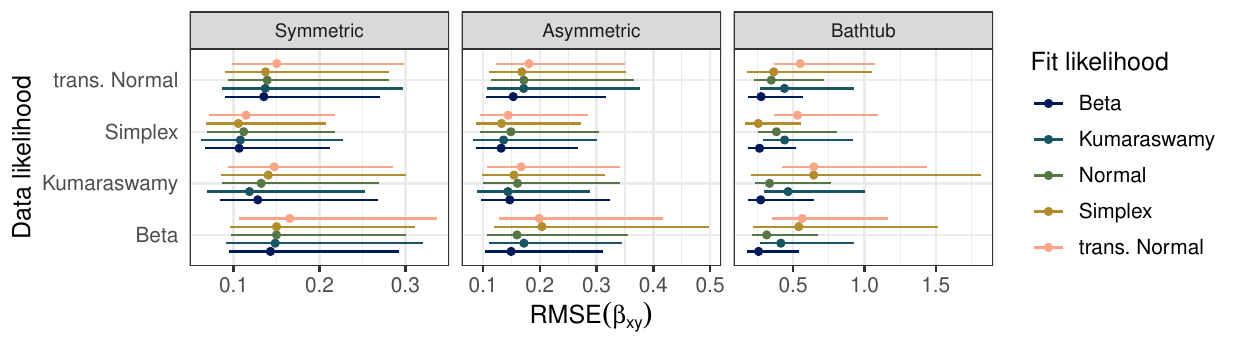}
\vspace{-10mm}
\caption{$\mathrm{RMSE}(\beta_{xy})$ performance for the double-bounded scenarios split by data generating likelihood and shape for the logit link. Only models with fit link = data link are included. The point indicates the median, while the range spans the central 95\% of the $\mathrm{RMSE}(\beta_{xy})$ values obtained in the simulations.}
\label{fig:unit-rmse_ll_logit}
\end{figure}

One of the most immediate observations is the stark difference between the bathtub shape on the one side and the symmetric and asymmetric likelihood shapes on the other side, as exemplarily shown in Figure \ref{fig:unit-rmse_ll_logit} for the $\mathrm{RMSE}(\beta_{xy})$ and logit link.
For both the symmetric and asymmetric likelihood shapes, the likelihood choice seems to have had little influence on $\mathrm{RMSE}(\beta_{xy})$ across all links.
For the bathtub shape, the beta likelihood showed the best average $\mathrm{RMSE}(\beta_{xy})$.
Depending on the scenario, the normal and Kumaraswamy likelihoods achieved similar performance but with less overall consistency.
The differences between links and data generating likelihoods were minor. As the only exception, we found that the cauchit-normal likelihood models performed considerably worse than all others.

\begin{figure}[tb]
\centering
\includegraphics[width=\linewidth]{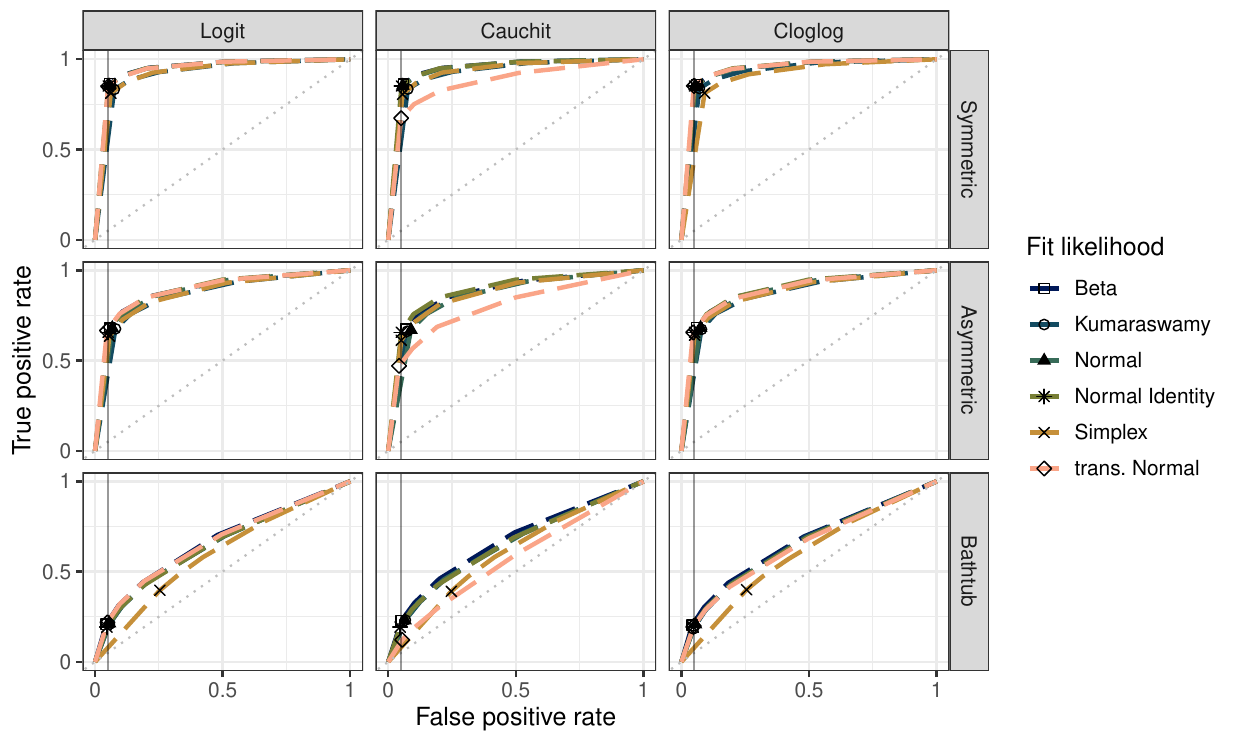}
\vspace{-10mm}
\caption{ROC for the double-bounded scenarios split by likelihood shape and fit links. The curves are calculated from the 50\%, 80\%, 90\%, and 95\% CIs, with the 95\% CI highlighted with a different symbol per fit likelihood. The vertical line indicates the nominal 0.05 FPR of the 95\% CI.}
\label{fig:unit-roc}
\end{figure}

In terms of error rates and ROC averaged over DGP link and likelihood family, as shown in Figure~\ref{fig:unit-roc}, there are two noteworthy observations.
First, likelihood choice again seems to have had little influence on calibration besides the worse performance of the cauchit-normal (middle column of Figure~\ref{fig:unit-roc}) and simplex (bottom row of Figure~\ref{fig:unit-roc}) likelihoods. This closely matches the patterns found for $\mathrm{RMSE}(\beta_{xy})$.
Second, the normal-identity (linear regression) model had error rates similar to the well performing canonical likelihood and link combinations.
The differences in error rates between data generating likelihoods and links as well as fit formulas were minor.

\begin{figure}[tb]
\centering
\includegraphics[width=\linewidth]{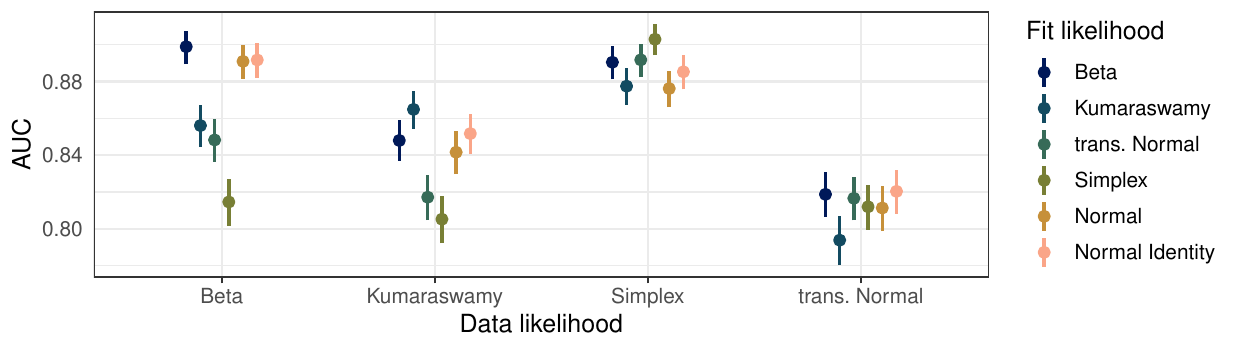}
\vspace{-10mm}
\caption{Conditional effects of data and fit likelihood on the area under the ROC curve (AUC) for the double-bounded scenarios.}
\label{fig:unit-auc-cond}
\end{figure}

Finally, Figure \ref{fig:unit-auc-cond} shows the conditional effects of the interaction of data and fit likelihood on the AUC. The results imply that the true likelihood family has the highest (median) AUC in each case, shortly followed by the normal likelihoods, both with the appropriate links and the identity link. Overall, we can observe very similar trends as with the other results, where differences are often small in absolute numbers. Depending on the application, the relative differences might however be relevant, as a false positive reduction from $0.05$ to $0.04$ would be a reduction of false-positives by $20\%$.
The conditional effects of the fit links on the AUC (shown in the online appendix \cite{online_appendix}) are very close to each other, with the logit link having the highest and the cauchit link having the lowest AUC for every data link.

\subsection{Lower-bounded Results}

\begin{figure}[tb]
\centering
\includegraphics[width=\linewidth]{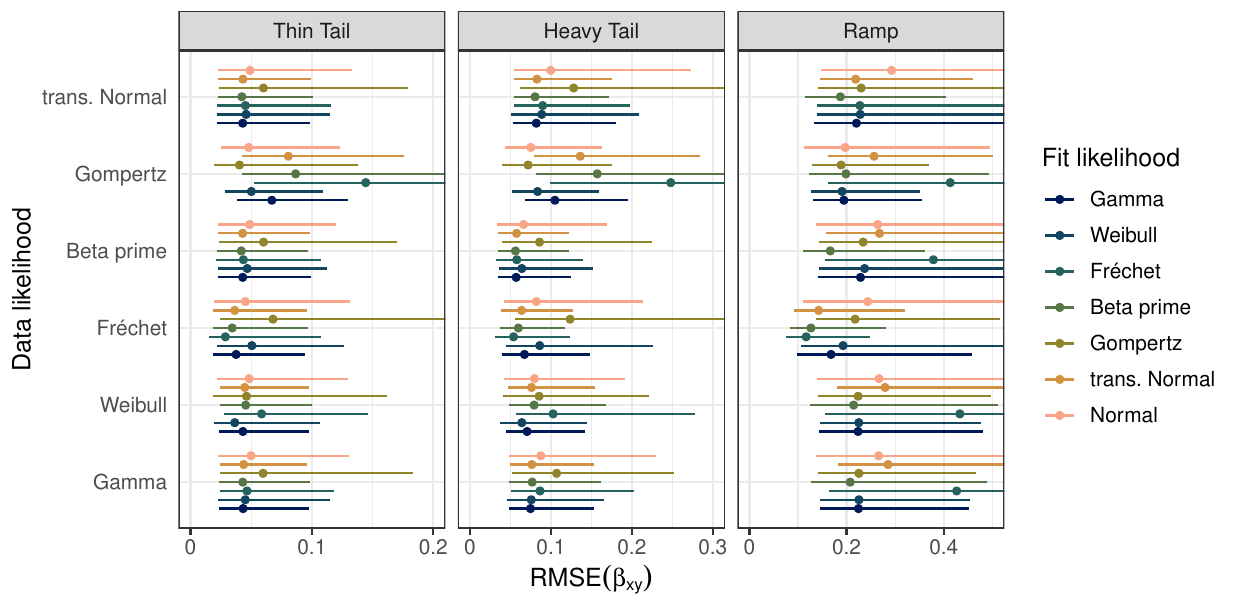}
\vspace{-10mm}
\caption{$\mathrm{RMSE}(\beta_{xy})$ performance for the lower-bounded scenarios split by data generating likelihood and shape for the log link. Only models with fit link = data link are included. The point indicates the median, while the range spans the central 95\% of the $\mathrm{RMSE}(\beta_{xy})$ values obtained in the simulations. The x-axes are truncated for improved readability.}
\label{fig:positive-rmse_log}
\end{figure}

Similar to the double-bounded data simulations, one of the most immediate observations is the difference between likelihood shapes. Specifically, for the log link, the ramp shape led to higher $\mathrm{RMSE}(\beta_{xy})$ compared to other likelihood shapes (see Figure~\ref{fig:positive-rmse_log}). In contrast, for the softplus link, the heavy tail shape implied the highest $\mathrm{RMSE}(\beta_{xy})$ (see the online appendix \citep{online_appendix}).
The differences between fit likelihoods were again small.
Notable exceptions to those general trends were the worse RMSE for the beta prime likelihood on all Gompertz data and softplus-normal heavy-tailed data. In addition, both the normal and transformed-normal likelihoods had worse recovery for the ramp shape compared to the other shapes.
The Gompertz and Fréchet likelihoods generally lacked consistency and had worse recovery than the other likelihoods outside of a few favourable scenarios.

In terms of error rates, we observed similar behaviour as for the double-bounded scenarios but with more variation between the fit likelihoods.
Similar to the results for $\mathrm{RMSE}(\beta_{xy})$, the Gompertz and Fréchet likelihoods showed an increased FPR across all scenarios, with less consistency for the remaining likelihoods in each individual case.
Similar to the double-bounded results, the normal-identity (linear regression) models had similar error rates to the well-performing but structurally faithful lower-bounded likelihood and link combinations.

\begin{figure}[tb]
\centering
\includegraphics[width=\linewidth]{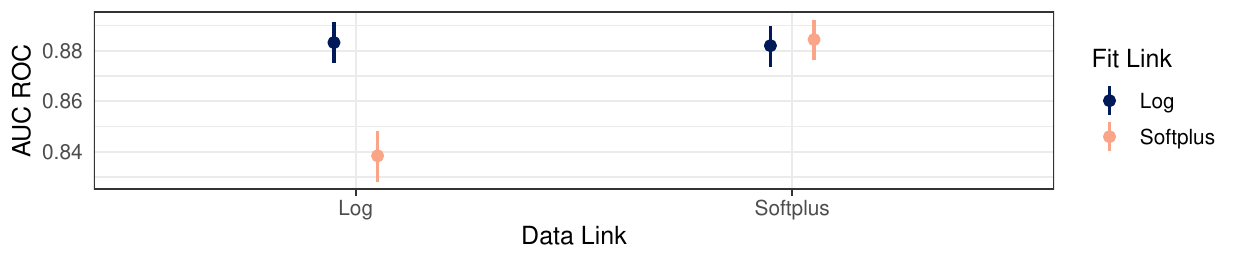}
\vspace{-10mm}
\caption{Conditional effects of data and fit link on the AUC for the lower-bounded scenarios. }
\label{fig:positive-auc-cond}
\end{figure}

For the lower-bounded scenarios, the conditional effects of the interaction between data and fit likelihood (not shown here) exhibited fewer differences among fit families compared to the double-bounded results.
Similar to the earlier results, the Fréchet and Gompertz distributions consistently had lower AUC as the others, while the differences between the latter were small. 
While the true likelihood family was always among the best performing ones, it is worth noting that it did not consistently yield a higher AUC than all alternative families.
The conditional effects of the interaction of data and fit link (see Figure \ref{fig:positive-auc-cond}) illustrate that the log link had consistently high (average) AUC for both log and softplus data, while the softplus link showed substantially lower AUC specifically for log data. At this point the exact mechanism that causes this asymmetry remains unclear.

\section{Discussion}
\label{sec:discussion}

This paper aims to advance our understanding of how the misspecification of likelihood families and link functions influences posterior accuracy and calibration of (Bayesian) GLMs. In this section we discuss the simulation results presented in Section~\ref{sec:results}, collect potentially useful learnings for practical applications, and provide suggestions for future research.
For brevity, we again use the term parameter recovery (PR) to refer to both posterior accuracy and calibration.


For both data types, we observed groups of well-behaving likelihoods that had similar PR. In both cases we also found a group of notably worse performing likelihoods, not only in terms of PR, but also in terms of sampling efficiency and convergence.
Accordingly, at least within the considered scenarios, likelihood choice appears to have little influence, as long as one chooses any of the well-behaving likelihood families.
Based on our results, we would advise practitioners to use a beta, Kumaraswamy or logit-normal likelihood for double-bounded data.
The beta distribution consistently led to good PR across all DGPs, while the Kumaraswamy and logit-normal likelihoods were slightly less consistent but still generally well-behaving.
For the lower-bounded data, we suggest the use of a log-normal, gamma, Weibull or beta prime likelihood. The log-normal likelihood showed the highest consistency and general good performance, with the gamma, Weibull and beta prime likelihoods again showing generally good performance but with less consistency.
Additionally, the normal likelihood was among the best performing likelihoods both when using with the matching link functions, for both data-types. However, care has to be taken here, as the normal likelihood does not respect the boundaries of the outcome data, such that we would only recommend its use in cases where this poses no risk.
The likelihoods that performed less consistently and overall worse generally had longer (fatter) tails, particularly the Fréchet and cauchit-normal likelihoods.
That being said, the respective true likelihood had among the lowest average $\mathrm{RMSE}(\beta_{xy})$ and $| \mathrm{bias}(\beta_{xy}) |$, as well as highest average AUC in all scenarios. The implication for practice is that one would still benefit from using the true likelihood, even if it were not part of the well-behaving group of likelihoods and PR is the main goal of the analysis. Section~\ref{sec:future-work} provides some examples of approaches for finding a well-fitting likelihood.

In terms of link function choice, the double-bounded links all produced similar results, though we found the logit link to generally be more stable during sampling and more commonly supported in statistics software packages. Here, we see little benefit of using alternative links based on our results and practical experience.
The lower-bounded links differed more strongly among each other, as the log link was clearly more robust than the softplus link.
These patterns certainly depend on the specifically included likelihood families and link functions and are likely subject to change if other link functions were included. This would imply the log link as the default choice for any analysis, however, we would prefer to better understand the underlying mechanism before a full endorsement.

In addition, we found that models using a normal likelihood with an identity link (i.e., linear regression) had both similar false-positive and true-positive rates to the structurally faithful alternatives.
This indicates that in practice, linear regression can be a valid alternative if frequentist calibration is the primary objective of an analysis, even when responses are bounded.
This result is also reassuring for the validity of many scientific results more generally as it shows that the practice of using linear regression can have good calibration even if it is not structurally faithful.
Additional relevance is provided by the fact that the estimation speed of linear regression models can be magnitudes faster than other GLMs, due do the availability of highly optimized implementations in both Bayesian and frequentist frameworks, an important aspect of especially Bayesian workflows, that often suffer from slow model fitting \cite{gelman_bayesian_2020, buerkner_utility_2022}.

All of the above recommendations could be considered \emph{default} options, i.e., they are already commonly used in reference and teaching materials and have good software support \citep{mcelreath_statistical_2020, gelman_bayesian_2013, stan_2022}.
To reiterate, if one is interested in posterior accuracy, we would suggest the above default choices as long as there is no additional information, e.g., from model comparisons \citep{scholz2023prediction}, that would support the use of a different likelihood or link: As exemplified by data generated from a simplex likelihood with bathtub shape, there definitely are cases where those \emph{default} likelihoods (or links) wouldn't be a good choice.
If calibration is the main objective of a statistical analysis, however, our results suggests that linear regression is as reliable as the structurally faithful likelihood and link alternatives, while potentially offering greatly increased computational performance due to optimized implementations. 

\subsection{Future Work}
\label{sec:future-work}

To our knowledge, the present paper is one of the largest comparison studies of its type with a total of over one million fitted models. Still, there are almost surely relevant likelihoods or link functions used in some research fields that we have missed.
A continuation of this work, focusing on the requirements of specific research fields could add valuable understanding in those areas.
Similarly, an extension to other data domains, such as unbounded-continuous, count, or ordinal data, would also help to assess the generalizability of our results.
This is especially relevant for the finding that linear regression (normal-identity) models achieve good error calibration across domains.

Finally, the use of a small set of likelihood shapes could have introduced a form of artificial similarity in the data generation process across likelihoods.
This could both favour likelihoods that accompany said shapes more easily and reduce differences between likelihoods by forcing them into the same shapes.
Future research in this direction could thus use a more general perspective on likelihood shapes, allowing to highlight differences among DGPs more clearly.
One option could be to sample the auxiliary parameters responsible for the likelihood shape from appropriate prior distributions to increase the diversity of true likelihood shapes.
Alternatively, one could use real data sets as the foundation of the data-generating process to more closely simulate problems encountered in practice.

\section*{Funding}
This work was partially funded by the Deutsche Forschungsgemeinschaft (DFG, German Research Foundation) under Germany’s Excellence Strategy -- EXC-2075 - 390740016 (the Stuttgart Cluster of Excellence SimTech).

This work was performed on the computational resource bwUniCluster funded by the Ministry of Science, Research and the Arts Baden-Württemberg and the Universities of the State of Baden-Württemberg, Germany, within the framework program bwHPC.

The authors gratefully acknowledge the support and funding.

\section*{Acknowledgements}
We also want to thank Marvin Schmitt and Daniel Habermann for their feedback on earlier versions of this manuscript as well as Yannick Dzubba for his help during development of the software used for our simulation study.

\section*{Data availability}
The data that support the findings of this study are openly available in OSF at \url{http://doi.org/10.17605/OSF.IO/TMDCF}.

\bibliographystyle{unsrtnat}
\bibliography{references}  

\end{document}